# A Tactile Sensor for Detection of Skin Surface Morphology and its Application in Telemedicine Systems

[1]Roozbeh Khodambashi, [1]Siamak Najarian, [2]Ali Tavakoli Golpaygani, [1]Alireza Keshtgar and [1]Shahla Torabi

[1]Artificial Tactile Sensing and Robotic Surgery Lab, Biomechanics Department, Faculty of Biomedical Engineering, Amirkabir University of Technology, Tehran, Iran
[2]Physical-Medicine Department, Shiraz University of Medical Sciences, Shiraz, Iran

**Abstract:** We proposed a new type of tactile sensor that is capable of determining the surface morphology of skin lesions. The sensor consists of a brass cylinder with an axial bore. Three peripheral bobbins were machined in the cylinder around which three coils have been wound. An iron core can easily move inside the bore. One of the coils acts as primary and the other two are secondary coils. Change in the position of the core due to surface peaks and valleys causes the induced voltage in the secondary coils to vary. This change in the voltage is a measure of core position and reveals the morphological features of the surface (surface profile). Experiments show that the proposed sensor is capable of discriminating between the models of two different types of skin lesions, namely, anetoderma and nodule, which resemble each other in visual inspection but behave differently when being touched. Thus, this sensor could be used as an assist device for diagnosis purposes in teledermatology systems.

## INTRODUCTION

Telemedicine is the use of information and communications technology to provide health care services to individuals who are some distance from the health care provider [1]. In the 1990s, serious interest in this method of health-care delivery has emerged. This is mainly due to improvements in the available technology, accompanied by falling transmission costs [2]. Many specialists ranging from image-dependent areas such as radiology and pathology to traditionally non-image-dependent areas such as psychiatry and surgery have found significant utility in telemedicine applications [3]. Although teledermatology is one of the best-studied disciplines in telemedicine, several of these issues remain unresolved and require further investigation [4]. At present, teledermatology is practiced in three different methods: store-and-forward (SAF), real-time (RT) or videoconferencing and combination of the two techniques. The SAF variant uses asynchronous data transfer technology (e.g., email) while RT teledermatology is based on synchronous data transfer technologies (e.g., videoconferencing software) [5]. Real-time video conferencing has been shown to be an effective substitute for the face-to-face consultation but is both time-consuming and expensive. With advent of the digital camera, store-and-forward systems with a history and digital image sent by email for an opinion, found to be accurate [6].

In addition to legal issues [7] and the problems related to acceptability of teledermatology systems [8,9], these systems are likely to have some technical problems. One of the major problems is the lack of tactile sense due to the available distance between the patient and the dermatologist. Dermatology is primarily a visual subject, but other sensory modalities, especially touch, and sometimes smell (e.g., anaerobic infection) are also important [10]. Skin surface morphology which reflects its surface characteristics such as softness, roughness, and the surface profile has a great importance in diagnosis of dermal diseases. Most of these properties are not recognized just by the sense of vision. To determine whether palpation alone could distinguish between two common dermatoses, a group of 16 patients were examined by a dermatologist using touch alone, with screens to prevent visualization of the lesions. The diagnosis was correctly made in 14 of 16 cases [10]. The most important problem mentioned by the physicians in the implementation of a calibrated

**Corresponding Author:** Prof. Siamak Najarian, No. 424, Hafez Avenue, Department of Biomechanics, Faculty of Biomedical Engineering, Amirkabir University of Technology, Tehran, Iran, P.O. Box 15875-4413. Tel: (+98-21)-6454-2378. Fax: (+98-21)-6646-8186

teledermatology system was the inability to feel the skin [11]. Despite of the importance of sense of touch in dermatology, present teledermatology systems are based solely on the photographs and films that are taken from the patient [11-13].

Several research activities have been performed on implementation artificial tactile sensing. Some of them have focused on estimating the physical properties of objects [14-16] and the others on increasing the diagnostic ability of medical instruments [17, 18]. Studies that are performed on skin surface properties include measurement of skin friction coefficient [19, 20] and skin topometry (for the study of the changes that take place due to lack of moisture, aging, and diseases) [21]. These methods are hard to be implemented in the clinics. Also, the structure of the devices used in them is such that only certain organs such as hands could be examined. Fabrication of a haptic finger for monitoring skin conditions have been studied [22]. This sensor, which could be easily placed on the fingertip, gathers required information as it is moved on the skin. The disadvantage of this sensor is that it cannot give exact information about skin's roughness and could only qualitatively compare different types of skin in their surface properties.

In this paper, fabrication of a tactile sensor will be considered which scans the surface of the skin and gives its surface profile. The size and structure of the proposed sensor is suitable to be used in most areas of human body. Also, the data gathered by this sensor could be used in teledermatology systems to help the dermatologist in making a faster and more reliable diagnosis.

**MATERIALS AND METHODS**

The tactile sensor system consists of a tactile sensor, a data acquisition device, and a personal computer for data analysis. The sensor is moved on the surface of the models of skin lesions that are made from paraffin gel. Data from the sensor is collected and sent to a PC where the surface profile of the lesion will be extracted. By comparing the surface profile of different lesions, a dermatologist can make a more reliable diagnosis.

**Tactile Sensor Structure**: The tactile sensor operation is based upon Faraday's induction law. According to this law, if a coil is placed in the varying magnetic field produced by another coil, a voltage will be induced in it which is proportional to the number of its turns and the rate of change of magnetic flux in it. This is given by:

$$\varepsilon = -N \frac{\partial \varphi}{\partial t} \quad (1)$$

In the above equation, $N$ is the number of turns of the coil, $\frac{\partial \varphi}{\partial t}$ is the rate of change of magnetic flux, and $\varepsilon$ is the induced voltage. The amount of flux passing the coil is related to the core material and the position of the core with respect to coil.

The tactile sensor is made up of one primary coil and two secondary coils which are shown by letters A and B in Fig. 1. The primary coil is excited by a sinusoidal voltage that generates a varying magnetic field which causes a voltage to be induced in each secondary coil. These voltages, which are shown by $V_a$ and $V_b$ in Fig. 1, are in phase but their magnitude is not equal and is calculated from Equation 1. Since the numbers of turns of the secondary coils are equal, the induced voltage is solely proportional to the position of the core. At the center (where an equal length of core is inside each coil), the induced voltages are equal. When the core moves up, the voltage of the secondary coil A ($V_a$) will become greater than the voltage of the secondary coil B ($V_b$) and the difference of these voltages ($V_a$-$V_b$) is positive in sign. When the core moves down, the sign of the voltage difference becomes negative. Thus, the magnitude of potential difference indicates the amount of motion and its sign indicates the direction of motion.

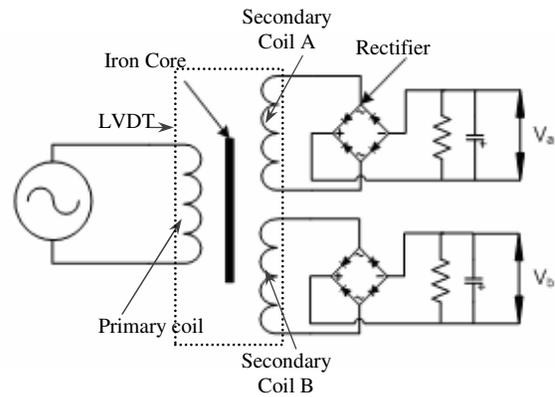

Fig. 1: A schematic of tactile sensor and rectifier circuit for detecting the direction of core movement

Fig. 2 shows different parts of the manufactured sensor. The core consists of three different pieces two of which are made of bronze and one is made of iron. The bronze pieces are attached to the ends of the iron core and act as holders. The body of the sensor is also made from bronze. The bronze is selected because its permeability is low and does not affect the flux that is passing through the coils. To decrease the friction between the core and the body, two linear bearings have been placed at the ends of the central hole.

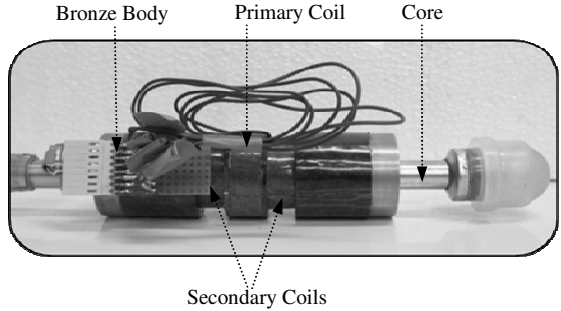

Fig. 2: Different parts of the manufactured sensor.

**Data Acquisition and Analysis System**: Since the output voltage of the sensor is small, it must be amplified before it can be sent to a computer for data processing. A data acquisition unit (PowerLab ML750 from AD Instrument) has been used for amplification and conversion of the analog signals to digital. The amplified signals are converted to digital and transmitted through the USB port to a PC where they are plotted versus time using the software package Chart5. Fig. 3 shows the required devices for acquiring, processing, and visualizing the data gathered by the sensor.

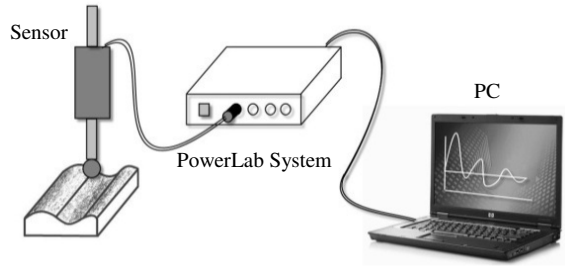

Fig. 3: Devices for acquiring, processing, and visualizing the information provided by the sensor

**Sensor Calibration**: For calibrating the sensor, the output voltage of the sensor has been obtained as a function of core displacement. As shown in Fig. 4 a micrometer has been used to move the core for 15 mm with steps of 0.02 mm. The sensor has been fixed using a hook. The core has been attached to the movable jaw of the micrometer and moves with it. The output voltage is recorded using the DAQ system.

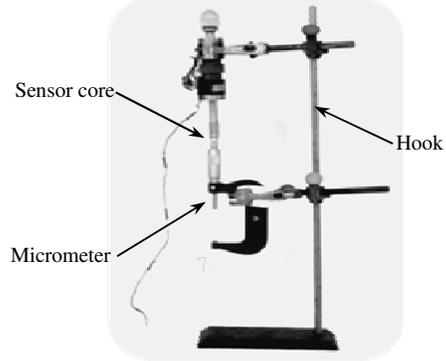

Fig. 4: Calibrating the sensor with a micrometer.

In Fig. 5, the output voltage of the sensor has been plotted as a function of core displacement. For a position sensor, linearity is defined as the percent of the full motion range for which this curve is a straight line. For our sensor, the curve could be considered as linear until the core moves 8 mm from the center. The full motion range is 15 mm and thus the linearity is 53.3%. In the linear range, the slope of the curve is 0.0145 V/mm and displacement of the core is obtained by:

$$d = \frac{V}{0.0145} \qquad (2)$$

In the above equation, $d$ is the displacement of the core and $V$ is the output voltage of the sensor.

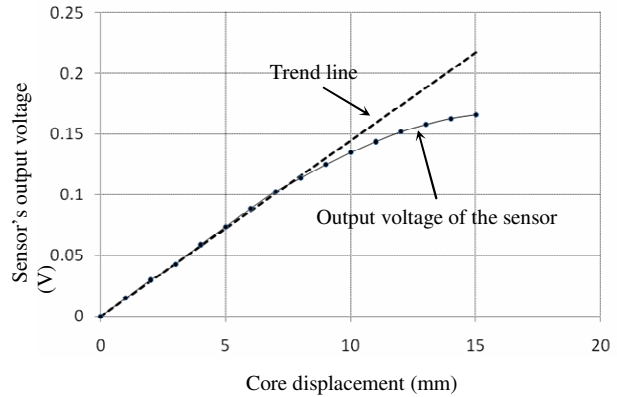

Fig. 5: The output voltage of the sensor as a function of core displacement

**Skin Surface Morphology:** The shape and distribution of surface rough is known as its morphology. Surface profile is a curve that is obtained from intersection of a surface with a hypothetical plain perpendicular to it. Surface profile is composed of several components namely roughness, waviness, and texture. Roughness is the smallest irregularities present in the surface which is usually felt through sense of touch and is not observed through vision. Waviness is the larger irregularities which could be seen by visual inspection and texture is the combination of roughness and waviness. Thus, the surface profile of an object is the sum of different functions with different wavelengths. Fig. 6 shows a sample surface profile and its components.

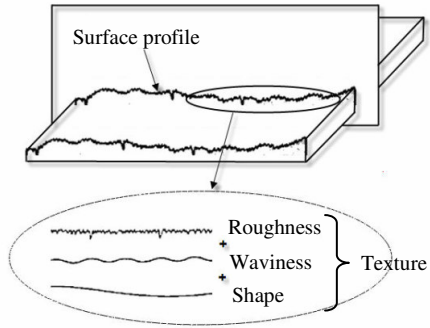

Fig. 6: A sample surface profile and its components.

Naturally, skin has some irregularities present in it which varies from one area of the body to another. Most of the skin diseases cause the shape of these irregularities to change. As a result, the surface profile of the skin differs from its normal condition. Fig. 7a shows a skin lesion which is known as anetoderma. The characteristic of this lesion is that it looks raised in visual inspection but unlike the nodule, its surface will be easily pushed down when being touched. This behavior is shown in Fig. 7b.

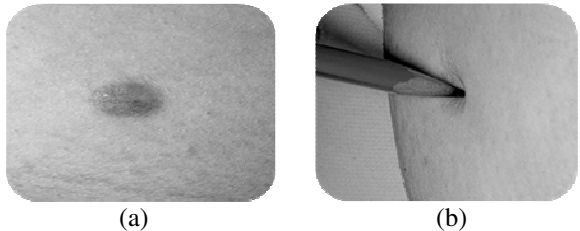

Fig. 7: Appearance of anetoderma, in (a) normal conditions and (b) when being touched

**Lesion Model Preparation:** To test the sensor's ability to distinguish between different skin lesions, artificial lesions have been prepared which their behavior is similar to the actual lesions. The model is shown in Fig. 8. This model is made up of a rigid cylindrical container which is blocked from one side and an elastic membrane is put on the other side. The pressure of the air inside the container is controlled through the pipe connected to it. This container is put in a larger rectangular container as shown in Fig. 8 and its surrounding is filled with paraffin gel which has mechanical properties similar to the skin. The sensor is moved with the aid of a robot from point 1 to point 2 in Fig. 8b in such a way that its tip remains in contact with the model surface.

The membrane is formed such that when the inside pressure of the container is equal to the pressure of surrounding air, it will stand as in position 1 in Fig. 8a. In this case, if the sensor tip is moved on it, it will deform to position 2. This behavior is similar to the behavior of anetoderma. If the pressure is increased above the surrounding pressure, the membrane will deform to position 3. In this case, it will deform to position 2 if the sensor's tip is moved on it.

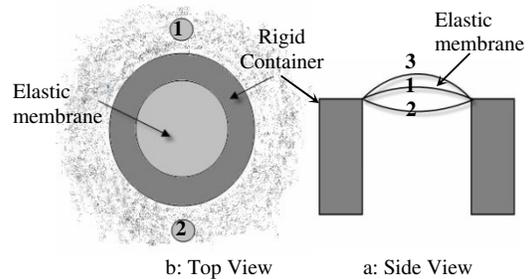

b: Top View      a: Side View
Fig. 8: Schematic of the skin lesion model.

### RESULTS AND DISCUSSION

In order to test the sensitivity of the sensor in detecting fine surface features, it was moved over a piece of paper with thickness of 100 microns. The sensor's output voltage is converted to displacement values using Equation 2 and is shown in Fig. 9. As can be seen, the sensor's output shows a sudden increase when it crosses the edge of paper. The thickness of the paper measured by the sensor is 110 microns which has an error of 10%.

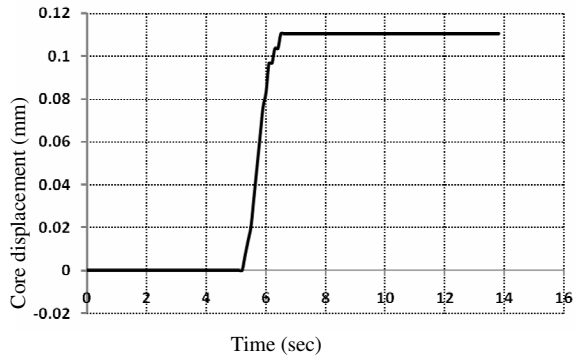

Fig. 9: Core displacement as a function of time while it crosses over a 100 microns thick paper.

The output voltage of the sensor when it passes over the model of Fig. 8 is converted to displacement values using Equation 2 and is plotted in Fig. 10. These curves are the surface profiles of the skin lesions. If the profile of the anetoderma is decomposed to its constituents, the curves of Fig. 11 and Fig. 12 will be obtained. Fig. 11 shows the overall shape of the surface of the model and indicates that point 1 in Fig. 8 is higher than point 2. Fig. 12 shows the waviness of the model surface which is formed due to presence of the lesion. The roughness of the surface could not be detected due to limitations of the sensor.

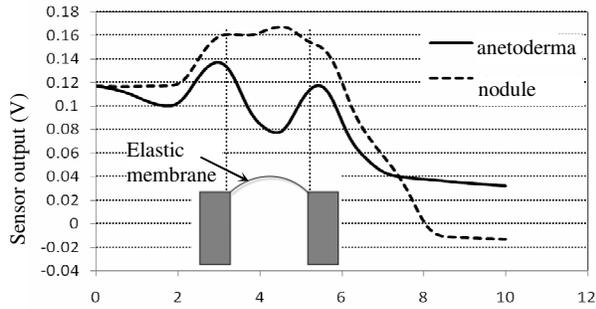

Fig. 10: Sensor's output voltage as it crosses over the model of anetoderma and nodule

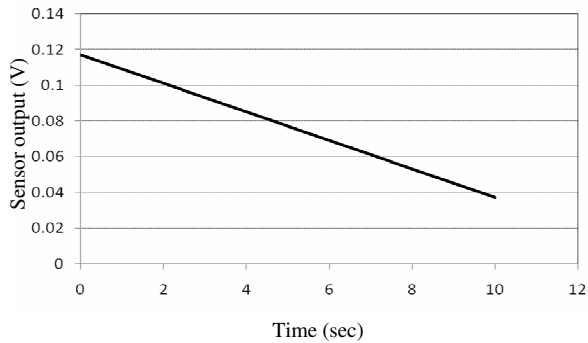

Fig. 11: Surface slope obtained from decomposition of anetoderma profile

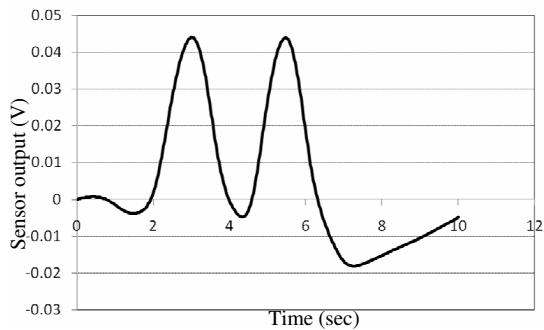

Fig. 12: Surface waviness obtained from decomposition of anetoderma profile.

## CONCLUSION

In this paper, fabrication of an inductive tactile sensor has been considered. This sensor is capable of quantitatively determining the surface profile of objects. The ability of this sensor in distinguishing between models of two skin lesions namely anetoderma and nodule justifies its use in teledermatology systems where it is difficult to make a correct diagnosis using visual inspections only.

Simplicity and ease of use are major features of this sensor and thus make it a suitable choice for use in medical clinics. Limitations of this sensor are its size and weight which makes it impossible to detect fine irregularities of soft skin tissue. This limitation could be overcome by making the sensor smaller through the use of advanced manufacturing techniques. Also, by using an array of these sensors, a 3-D surface profile of skin could be obtained.

## ACKNOWLEDGEMENTS

The authors would like to express their gratitude to the Center of Excellence of Biomedical Engineering of Iran based in Amirkabir University of Technology, Faculty of Biomedical Engineering for its contribution.